# Electrically Detected Paramagnetic Resonance in Ag-paint Coated Polycrystalline DPPH

Lee Yong Heng, Ushnish Chaudhuri, and Ramanathan Mahendiran [*]

Department of Physics, National University of Singapore, 2 Science Drive 3,

Singapore 117551, Singapore

**Abstract**

We describe a simple experimental method to detect electron paramagnetic resonance (EPR) in polycrystalline 2,2-diphenyl-1-picrylhydrazyl **(**DPPH) sample, the standard *g*-marker for EPR spectroscopy, without using a cavity resonator or a prefabricated waveguide. It is shown that microwave (MW) current injected into a layer of silver paint coated on an insulating DPPH sample is able to excite the paramagnetic resonance in DPPH. As the applied dc magnetic field $H$ is swept, the high-frequency resistance of the Ag-paint layer, measured at room temperature with a single port impedance analyzer in the MW frequency range $f$ = 1 to 2.5 GHz, exhibits a sharp peak at a critical value of the dc field ($H = H_r$) while the reactance exhibits a dispersion-like behavior around the same field value for a given $f$. $H_r$ increases linearly with $f$. We interpret the observed features in the impedance to EPR in DPPH driven by the Oersted magnetic field arising from the MW current in the Ag-paint layer. We also confirm the occurrence of EPR in DPPH independently using a co-planar waveguide based broadband technique in the frequency range 2 – 4 GHz. This technique has the potential to investigate other EPR active inorganic and organic compounds.

[a] Physics Department, 2 Science Drive 3, National University of Singapore, Singapore 117551, Republic of Singapore

[*]Author for correspondence (phyrm@nus.edu.sg)

## Introduction

Electron paramagnetic resonance spectroscopy (EPR) is an invaluable experimental tool for physical chemists to study interactions of unpaired electrons with their surroundings to understand local environment, spin states, hyperfine interaction with nuclear magnetic moments[1] and finds diverse applications such as in studying structural phase transition in perovskite oxides[2], distinguishing localized and delocalized electronic states in N doped 6H SiC[3], redox process in metallization of Li in battery[4], etc. The DPPH (2,2-diphenyl-1-picrylhydrazyl, $C_{18}H_{12}N_5O_6$) molecule is a stable free radical solid source with a narrow EPR linewidth, and it has been used as a standard position and intensity marker (*g*-marker) for EPR spectrometers for a long time [5]. It possesses one unpaired spin per 41 atoms and the solvent free DPPH is paramagnetic above 15 K but its magnetic susceptibility shows a maximum around 11 K due to spin paring at lower temperatures. [6,7] When a DPPH sample, placed inside a microwave (MW) cavity resonator, is simultaneously exposed to a mutually perpendicular electromagnetic (EM) radiation and slowly varying dc magnetic field, the unpaired electrons from the Zeeman-split lower energy level will be transferred to higher energy level by resonantly absorbing energy from the MW irradiation. In conventional EPR spectrometer, the output signal is usually in terms of the field derivative of power absorption ($dP/dH$) which undergoes zero crossing at the resonance field ($H = H_r$). DPPH exhibits a single spectral line in X-band (~9 GHz) with a small linewidth, i.e.: ~1.5–4.7 Oe.[5-7] Conventional EPR spectrometers are bulky and operate at a single resonant frequency in X-band ($f \approx 9.5$ GHz) which is determined by the size and shape of the resonant cavity. Multiple microwave cavities of different sizes must be used for measurements at different frequencies. Broadband spectrometers are being developed that take advantage of powerful vector network analyzers to source MW signal over a wide range of frequencies (~10 MHz – 40 GHz) and detect variation in the transmitted signal due to MW absorption by a sample placed on a coplanar waveguide[8,9] or a microstrip resonator. [10,11,] or a transmission line.[12] Planar microcoil[13], dual antenna[14], copper strip coil[15]and surface loop-gap micro resonator[16] were also exploited to inductively detect EPR in DPPH molecule.

Here, we demonstrate a novel electrical method to detect EPR. It is possible to electrically excite unpaired electrons and detect paramagnetic resonance in a polycrystalline DPPH sample without using a waveguide or a microcoil but through passing MW current in a layer of silver paint coated on polycrystalline DPPH surface and measuring the high-frequency electrical impedance of the Ag-paint using an impedance analyzer. We believe that the simplicity of this method will be helpful to investigate spin resonance and relaxation in other EPR active samples that are difficult to be synthesized as single crystals or fabricated as thin films.

**Experimental method**

The DPPH powder was procured from a commercial chemical supplier (Sigma-Aldrich Co.). Since it has a low melting point (~ 135°C), we cold-compressed the DPPH powder into a pellet of a 10 mm diameter disc by applying uniaxial pressure of 6 tons/inch$^2$ using a hydraulic press. A rectangular sample of dimensions 4 mm x 2mm x 2 mm was cut from the disc for magnetoimpedance measurement. The two probes resistance of DPPH was the order of $10^9$ ohm at room temperature and hence it was not possible to measure its magnetoimpedance by coating silver paint only at two ends of the sample as done in the case of Mn-perovskites whose resistivities of the order of a few milli ohm cm at room temperature.[17,18] Hence, in addition to coating both ends of the sample, the top surface of the sample was also coated with silver paint and was air dried at 80$^o$C for 10 minutes. Then, the sample was loaded in a sample probe in which the sample bridges the signal line and the ground plane. The sample holder was placed at the center of an electromagnet. A schematic sketch of our experimental setup is illustrated in Fig. 1(a). An Agilent E4991 RF impedance analyzer was used to measure the electrical impedance of the sample while sweeping dc magnetic field provided by the electromagnet. The output power of the impedance analyzer was set to 0 dB (= 1 mW) so that the sample does not heat up and the signal is in the linear regime. This impedance analyzer uses a single port to source the microwave current that flows from the signal line to ground plane via the sample and returns to the port of the instrument. The instrument calculates the impedance by measuring current vs voltage for a given frequency. A graphical software (Labview™) allows us to record the resistance (*R*) and reactance (*X*) of the sample simultaneously while sweeping the frequency of microwave from 0.9 GHz to 3 GHz for each selected value of the magnetic field while the dc magnetic field was swept slowly.

To verify the results obtained by the magnetoimpedance measurement, we have also investigated microwave absorption in the same sample (without silver paint) using a home-made broadband spectrometer, shown schematically in Fig. 1(b). A microwave signal generator (Anirutsu MG 3693C) was used to inject the microwave signal into a coplanar waveguide (CPW). The CPW, with the sample on it, was placed at the center of two pole-pieces of an electromagnet. Due to the limitation of the maximum field achievable with the 12 mm pole-to-pole spacing, the maximum field in our case is limited to $H$ = 1.4 kOe. A pair of Helmholtz coils were used to modulate the dc magnetic field at a frequency of 178.3 Hz and a rf diode (Schottky diode PE8013) was used to rectify the transmitted microwave signal from the CPW at 178.3 Hz and detected as a dc voltage by a lock-in amplifier (SRS model 830). The signals are measured in the first derivative of power, i.e., as $dP/dH$.

**Results and Discussions**

Figure 2(a) shows the magnetic field dependence of *R* and *X* responses for $f$ = 2 GHz. We only show the data for a limited field range for clarity. As $H$ is swept from 900 Oe to 550 Oe, *R* shows a sharp peak at 735 Oe while *X* displays a rapid increase at the same field value which is immediately preceded by a minimum and followed by a maximum. The shapes of $R(H)$ and $X(H)$ resemble absorption and dispersion curves predicated for the out-of-phase ($\chi''$) and the in-phase ($\chi'$) components of the complex susceptibility ($\chi$= $\chi'$-i$\chi''$) respectively.[19] The power absorption measured in commercial EPR spectrometer is proportional to $\chi''$. Figure 2(b) shows the *R* and *X* responses as *f* increases from 1 GHz to 2.5 GHz. As the sample showed dielectric resonance around 2.6 GHz, we have to confine our data to a lower range of *f*. As *f* increases, the absorption and dispersive signals clearly shift to a higher field value, i.e., 400 Oe at 1 GHz to ~1 kOe at 2.5 GHz. While a similar shift of the peak position in *R* with increasing frequency was also noted previously in manganites[17,18] the peak in *R* is distinctively sharp and the changes in *X* are more dramatic in DPPH as compared to the broad features observed in manganites.

The signals in Figure 2(b) were analyzed by fitting to Eq. 1, a set of Lorentz functions with a symmetric Lorentzian term describing the absorption and an asymmetric term describing the dispersion.

$$R\ (or\ X) = A_{sym}\frac{(\Delta H)^2}{(H-H_r)^2+(\Delta H)^2} + A_{asym}\frac{(\Delta H)(H-H_r)}{(H-H_r)^2+(\Delta H)^2} + C \qquad (1)$$

$\Delta H$ and $H_r$ are the linewidth and resonance field for a particular frequency. $A_{sym}$ and $A_{asym}$ reflect the maximum amplitudes of the absorptive and dispersive components, and *C* is an offset constant. The Lorentz fits using Eq. 1, were performed onto the *R* signals at 5 different frequencies, as shown in Figure 3(a). The red solid lines depict the fits, and we can note that the fit closely matches to the data signals. Similar procedure was performed for all other frequencies and the line shape analysis were carried out to extract the frequency dependent linewidths ($\Delta H$) as well as the resonance fields ($H_r$). The plot $H_r$ versus frequency of the current plotted in figure 3(b) reveals that $H_r$ roughly increases linearly with the frequency of the MW current. A linear dependence of $H_r$ versus frequency of microwave radiation is expected for EPR. The resonance condition for EPR is given by[1]:

$$f = (\gamma/2\pi)H_r \qquad (2)$$

where $\gamma$ is the gyromagnetic ratio ($\gamma = g\mu_B/\hbar$, where $g$ is the Landé $g$-factor, $\mu_B$ is the Bohr magneton and $\hbar$ is the reduced Plank's constant). The experimental data are shown by blue solid circles and the red solid line is the fit to the above EPR relation. The slope of the fit yields $\gamma/2\pi = 2.769 \pm 0.003$ MHz/Oe and from this we obtain a Landé $g$-factor = $1.978 \pm 0.002$ which is very close to the literature value of 2.0036 for DPPH arising from spin-only contribution. The linewidth is within 2 Oe as shown in figure 3(c). This is consistent with the dilute nature of paramagnetic species[12].

A likely origin of the peak in $R$ is the occurrence of the paramagnetic resonance in DPPH. The flow of MW current in the Ag-paint generates a small oscillating MW magnetic field (Oersted field) $h_{MW}$ which acts on spins of free radicals in DPPH. If the applied dc magnetic field and the $h_{MW}$ are perpendicular to each other, spins of free radicals in DPPH can resonantly absorb energy from the MW magnetic field to flip spin directions from antiparallel to parallel to dc magnetic field when the energy of the microwave magnetic field matches with the Zeeman splitting. While the MW current in manganites can pass directly through the sample because of their low resistivities, MW current in insulating DDPH flows mostly in the Ag-paint layer yet it is able to induce spin resonance in the DPPH layer beneath. We are able to detect the magnetization dynamics in the insulating DPPH through a sudden change in the resistance of the Ag-paint itself. We verified that Ag-paint coated on a glass plate does not show magnetoresistance and the observed $R$ peak in Ag-paint is indeed caused by the magnetic resonance in DPPH. We assume that Ag-paint acts as waveguide for MW current, which induces MW magnetic field that interact with spins of the DPPH. Since the resistivity of DPPH is in the order of Giga ohm cm, the MW magnetic field can penetrate fully into the whole volume of the sample unlike in manganites. We can apply the results of Greenberg and Koch who proposed an electrical detection method for nuclear magnetic resonance in metals using superconducting quantum interference devices.[20] By simultaneously solving Bloch's equation for magnetic relaxation and Maxwell's equation for electromagnetic wave propagation, they obtained expression for changes in high frequency resistance and inductance for the case of a wire and a thin film.

Suppose if a sinusoidal alternating current of angular frequency ω flows in the thin film of thickness $2a$ and width $d$ and length $l$, the surface impedance of the thin film is

$$Z = R(ka)\cosh(ka)/\sinh(ka) \tag{3}$$

where $R$ is the dc resistance of the film and $k$ is the wave vector $\left(k^2 = i\frac{2}{\delta^2}[1+\chi(\omega)]\right)$ related to skin depth $\delta$ given by $\delta = \left(\frac{2\rho}{\omega\mu_0}\right)$ where $\rho$ is the dc resistivity and $\mu_0$ is the permeability of the vacuum.

The impedance in the low frequency limit ($a/\delta$ <<1) in the fourth order of $a/\delta$ can be written as

$$Z = [R + \Delta R] + i\omega[L + \Delta L] \tag{4}$$

where $\Delta R$ and $\Delta L$ are the excess resistance and inductance due to resonance, and they are given by

$$\Delta R = \mu_0 \omega l \left(-\frac{a}{6d}\right)\left[\chi''(\omega) - \frac{2a^2}{15\delta^2}\right] \tag{5}$$

$$\Delta L = \mu_0 l \left(\frac{a}{6d}\right)\left[\chi'(\omega) + \frac{4a^2}{15\delta^2}\chi''(\omega)\right] \tag{6}$$

Since $a/\delta$ << 1 for our sample, the 2$^{nd}$ terms in $\Delta R$ and $\Delta L$ are much smaller than the 1$^{st}$ terms and can be neglected. If the width of the sample ($d$) is taken as twice the thickness (2$d$), then we can write

$$|\Delta R| \cong \frac{\mu_0 \omega l}{24}\chi''(\omega) \tag{7}$$

$$\Delta L \cong \frac{\mu_0 l}{24}\chi'(\omega) \tag{8}$$

Hence, the excess resistance of the sample is dependent on the out of phase component $\chi''(\omega)$ of the susceptibility. Theory predicts that $\chi''(H)$ exhibits a peak the $\chi'(H)$ for a single frequency will cross over zero at the critical value of dc magnetic field when spins of DPPH undergo paramagnetic resonance and the reactance will cross over zero at the resonance. Recently, Sing and Rani proposed an alternative scheme for the electrical detection of EPR involving momentum randomization of Zeeman split electrons via spin fluctuations.[21]

To convince ourselves that the peak observed in the field dependence of $R$ is caused by paramagnetic resonance in the DPPH sample, we measured the microwave power absorption in DPPH using a home-built setup in which the DPPH sample, placed on a CPW, experiences microwave magnetic field generated by MW current in CPW. The experimental set up records the spectrum as the field derivative power absorption (*dP/dH*) as the field is decreased from the maximum field (H = -1.4 kOe) to H = 0 kOe. Results are shown in Fig. 4 (a), where the spectra for different frequencies are displaced along the vertical scale. As the field is decreased from the maximum negative field to zero, the spectrum at any frequency shows a dip at a higher field followed by a peak at a lower field. We can see that the spectra shift towards higher magnetic field with increasing frequency of the MW signal. The resonance field is where the *dP/dH* curve cross over the zero value. Similar to the *MWMR* data, the *dP/dH* curves were fitted to a sum of absorption and dispersion term as follows:

$$\frac{dP}{dH} = A_{asym}\frac{4\Delta H(H-H_r)}{[4(H-H_r)^2+(\Delta H)^2]^2} - A_{sym}\frac{(\Delta H)^2-4(H-H_r)^2}{[4(H-H_r)^2+(\Delta H)^2]^2} + C \qquad (9)$$

By fitting into Eq. 6, we determine the resonance field $H_r$. Figure 4(b) shows that the resonance field $H_r$ increases linearly with increasing frequency. From the fit, we obtained $\gamma/2\pi$ = 2.83 $\pm$ 0.003 MHz/Oe with $g$ = 2.022 $\pm$ 0.002, which agree with the values obtained in our MI measurement.

There is no previous report on the detection of EPR of DPPH through measuring magnetoresistance of the Ag-paint layer. However, after the completion of the present work, we came across a related study of MW current-driven ferromagnetic resonance (FMR) in YIG/Pt bilayer thin films.[22] YIG is a ferrimagnetic insulator unlike the paramagnetic DPPH and Pt is a non-magnetic metal having strong spin-orbit coupling. The YIG/Pt bilayer was integrated into a CPW by dissecting the signal line to insert the bilayer. Passing the microwave current in the Pt layer excited FMR in YIG and it was detected as a peak in dc voltage in the Pt layer. It was suggested that apart from the Oersted field induced torque, another torque known as the spin-transfer torque also acted on the spins at the interface between YIG and Pt. High-frequency current in the Pt layer generates an ac spin current owing to its spin-orbit interaction in the Pt layer and it is injected into the YIG layer. The ac spin current exerts an ac spin-transfer torque on the magnetization of the YIG and drives it to FMR. The spin-transfer torque can couple the magnetization orientation of the YIG to the electrical resistance of the Pt film via spin Hall magnetoresistance. As a result, the electrical resistivity of Pt film is modulated at the same frequency of the high-frequency current and spin rectified in the Pt layer through anisotropic magnetoresistance and detected as a dc voltage. Angle dependence of the FMR and dc voltage led to the conclusion that while the Oersted field effect was the dominant source of FMR for thick YIG films (~50 nm), spin-transfer torque dominated FMR in thinner (~4 nm) films. Our sample is 2 mm thick polycrystalline paramagnet. Our detection method is based ac magnetoresistance/magnetoreactance, instead of a dc voltage, of the non-magnetic layer Ag-paint which is made up colloidal Ag nanoparticles in a binder. Spin-orbit coupling is weak in Ag compared to Pt. While we interpret our results as the Oersted-field induced electron paramagnetic resonance driven by the MW current, it is worth to study thin films of DPPH with different thickness and sputtered silver or platinum layer instead of silver paint to disentangle different contributions to the observed signal.

**Conclusion**

In summary, we have shown that microwave magnetoresistance of Ag-paint coated on paramagnetic DPPH sample, bears imprints of paramagnetic resonance of the DPPH sample. We interpret the results as the microwave current induced Oersted field driving the paramagnetic resonance in DPPH. This experimental method can be exploited to characterize other polycrystalline materials which have EPR active species. This technique will be more attractive and beneficial if temperature dependence measurements can be incorporated. Also, to disentangle different contributions of spin torques, thin films of bilayer DPPH/NM with different nonmagnetic metals (NM= Pt, Ta, Au, Ag) have to be investigated in future.

**Acknowledgements**

R.M. acknowledges the Ministry of Education, Singapore (grant number R144-000-442-114).

**Conflicts of interest:**

The authors declare no conflict of interest.

**Data availability statement:**

The data used in the present work will be available to reader on reasonable request.

**References**

(1) C. P. Poole, Jr. Electron spin resonance: A comprehensive Treatise on experimental Techniques. 2nd Edition, John-Wiley and Sons, New York, 1983.

(2) K. A. Muller; W. Berlinger; F. Waldner. Characteristic structural phase transition in perovskite-type compounds. *Phys. Rev. Lett.* **1968**, 21, 814.

(3) D. Savchenko; E. Kalabukhova; B. Shanina; S. Cichon; J. Honolka; V. Kiselov; E. Mokohov. Temperature dependent behavior of localized and delocalized electrons in nitrogen-doped 6H SiC crystals as studied by electron spin resonance. *J. Appl. Phys.* **2016**, 119, 045701.

(4) J. Wandt; C. Marino; H. A. Gasteiger; P. Jakes; R-A. Eichel; J. Granwehr. Operando electron paramagnetic resonance spectroscopy – formation of mossy lithium on lithium anodes during charge-discharge cycling. *Energy Environ. Sci*. **2015**, 8, 1358.

(5) H. Ueda; Z. Kuri; S. Shida. Electron-Spin-Resonance Studies of DPPH Solutions. *J. Chem. Phys*. **1962**, 36, 1676.

(6) T. Fujito. Magnetic Interaction in Solvent free DPPH and DPPH-Solvent Complexes. *Bull Chem. Soc. Japan*, **1983**, 54, 3110.

(7) P. Grobet; L. Van Gervin; A. Van den Bosch. The spin magnetism of α, α′-diphenyl-β-picrylhydrazyl (DPPH). *J. Chem. Phys*. **1978**, 68, 5225.

(8) Y. Wiemann; J. Simmendinger; C. Clauss; L. Bogani; D. Bothner; D. Koelle; R. Kleiner; M. Dressel; M. Scheffler. Observing electron spin resonance between 0.1 and 67 GHz at temperatures between 50 mK and 300 K using broadband metallic coplanar waveguides. *Appl. Phys. Lett*. **2015**, 106, 193505.

(9) K. Jing; Z. Lan; Z. Shi; S. Mu; X. Qin; X. Rong; J. Du. Broadband electron paramagnetic resonance spectrometer from 1 to 15 GHz using metallic coplanar waveguide. *Rev. Sci. Instrum.* **2019**, 90, 125109.

(10) B. Johansson; S. Haraldson; L. Pettersson; O. Beckman. A stripline resonator for ESR. *Rev. Sci. Instrum*. **1974**, 45, 1445.

(11) A. C. Torrezan; T. P. Mayer Alegre; G. Medeiros-Ribeiro. Microstrip resonators for electron paramagnetic resonance experiments. *Rev. Sci. Instrum*. **2009**, 80, 075111.

(12) W. R. Hagen. Broadband Tunable Electron Paramagnetic Resonance Spectroscopy of Dilute Metal Complexes. *J. Phys. Chem. A*. **2019**, 123, 6986.

(13) G. Boero; M. Bouterfas; C. Massin; F. Vincent; P.-A. Besse; R. S. Popovic. Electron-spin resonance probes based on 100 mm planar microcoil. *Rev. Sci. Instrum*. **2003**, 74, 4794.

(14) Z. H. Jang; B. J. Suh; M. Corti; L. Cattaneo; D. Hajny; F. Borsa; M. Luban. Broadband electron spin resonance at low frequency without resonance cavity. *Rev. Sci. Instrum*. **2008**, 79, 046101.

(15) U. Chaudhuri; R. Mahendiran. Detection of L-band electron paramagnetic resonance in DPPH molecule using impedance measurements. *RSC Advances* **2020**, 10, 17311.

(16) Y. Twig; E. Suhovoy; A. Blank. Sensitive surface loop-gap microresonators for electron spin resonance. *Rev. Sci. Instrum*. **2010**, 81, 104703.

(17) A. Chanda; U. Chaudhuri; R. Mahendiran. Broadband magnetotransport in $La_{0.6}Sr_{0.4}Mn_{1-x}Ga_xO_3$. *J. Appl. Phys*. **2019**, 126, 083905.

(18) A. Chanda; R. Mahendiran. Current-Driven Spin Precession in $Nd_{0.6}Sr_{0.4}MnO_3$. *J. Phys. Chem. C*. **2020**, 124, 18226

(19) A. G. Marshall; D. C. Roe. Dispersion versus absorption: spectral line shape analysis for radiofrequency and microwave spectrometry. *Anal. Chem*. **1978**, 50, 756.

(20) Y. S. Greenberg; H. Koch. The excitation of NMR transitions by the current in a sample and the proposal for its detection. *Solid State Nucl. Magn Reason.* **1998**, 11, 129.

(21) N. Singh; L. Rani. A novel effect of Electron Spin Resonance on electrical resistivity. *J. Magn. Magn. Mater.* **2019**, 474, 501.

(22) M. Schreier; T. Chiba; A. Niedermayr; J. Lotze; H. Huebl; S. Geprags; S. Takahashi; G. E.W. Bauer, R. Gross; S. T. B. Goennenwein. Current-induced spin torque resonance of a magnetic insulator. *Phys. Rev. B.* **2015**, 92, 144411.

## Figures and Captions

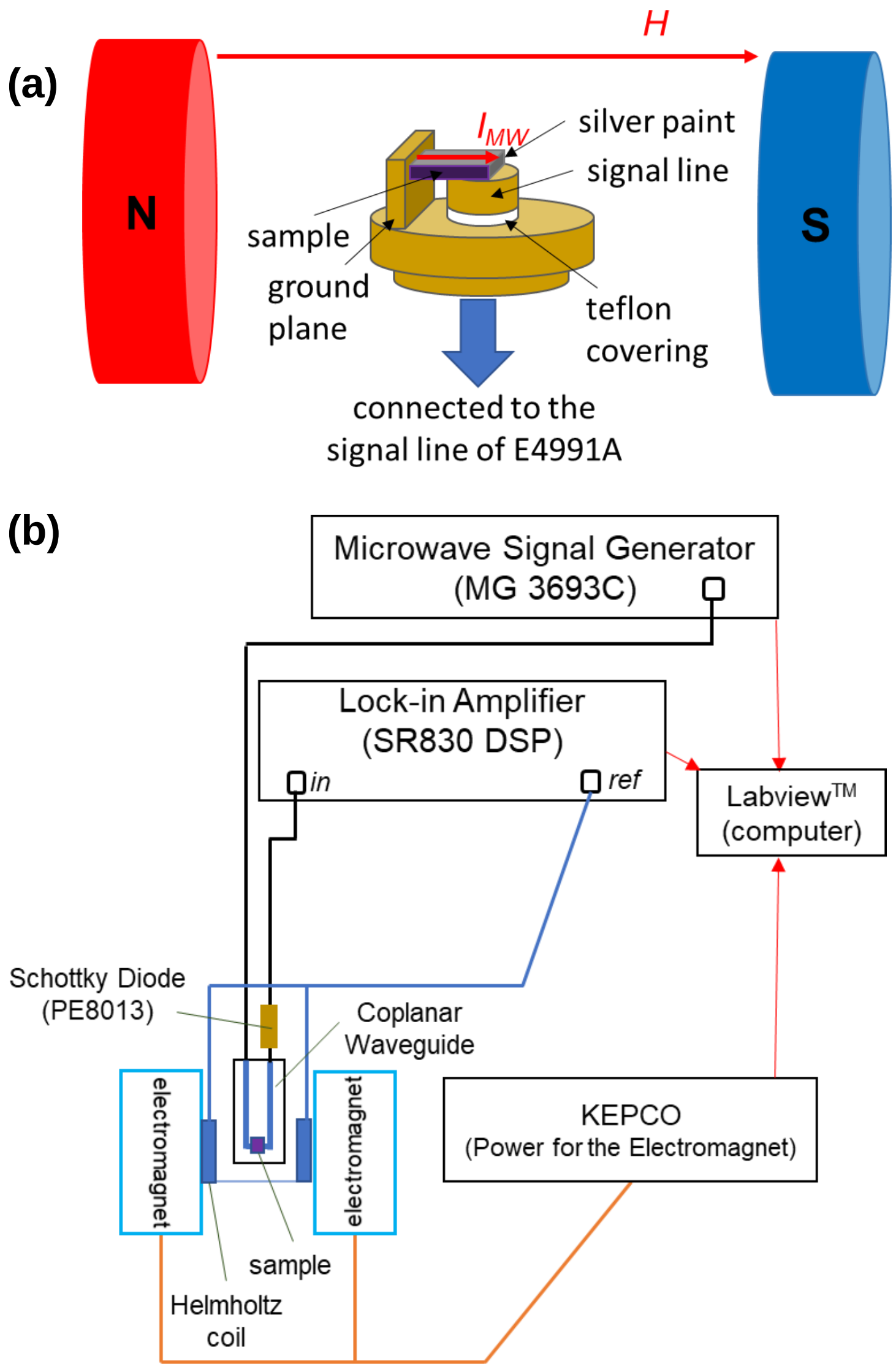

**Figure 1**. (a) A schematic diagram of the DPPH sample mounted on the sample probe using the silver-paste, with the direct of microwave current flow in the sample $I_{MW}$ that generates microwave magnetic field $h_{MW}$ in the sample perpendicular to the applied dc magnetic field $H$ (b) A schematic diagram of the broadband magnetic resonance spectrometer in which the sample is placed on a coplanar wave guide. connected to a lock-in Amplifier and the microwave signal generator with $h_{MW}$ perpendicular to $H$.

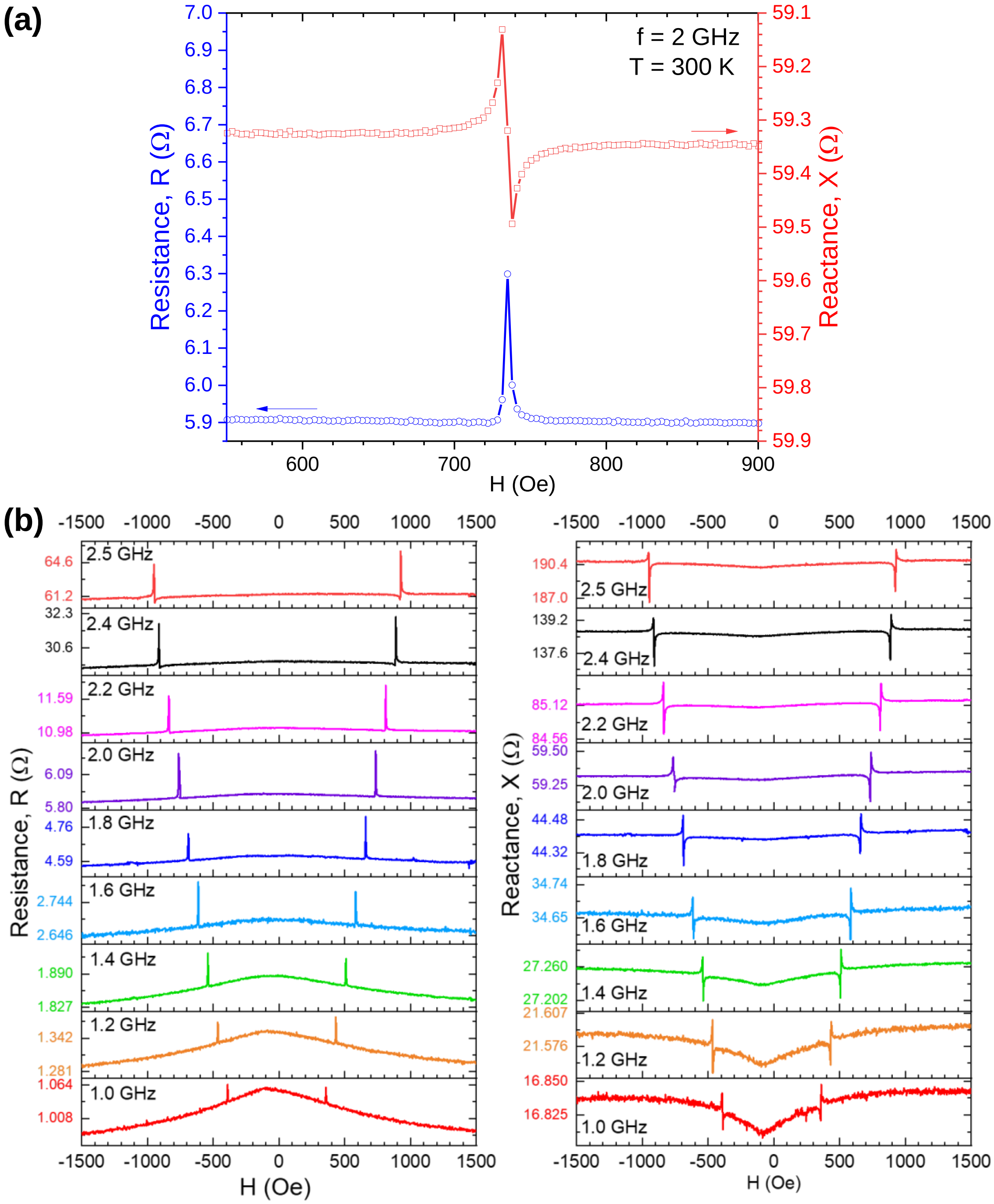

**Figure 2.** (a) Magnetic field dependence of Resistance (*R*) and reactance (*X*) of the DPPH-Ag sample measured at room temperature with excitation frequency $f$ = 2 GHz. The prominent features seen in both the *R* and *X* data at 735 Oe are caused by electron paramagnetic resonance in DPPH. (b) *R*(*H*) and *X*(*H*) at selected frequencies, from $f$ = 1 to 2.5 GHz. The prominent features observed in *R* and *X* move toward higher magnetic fields with increasing frequencies.

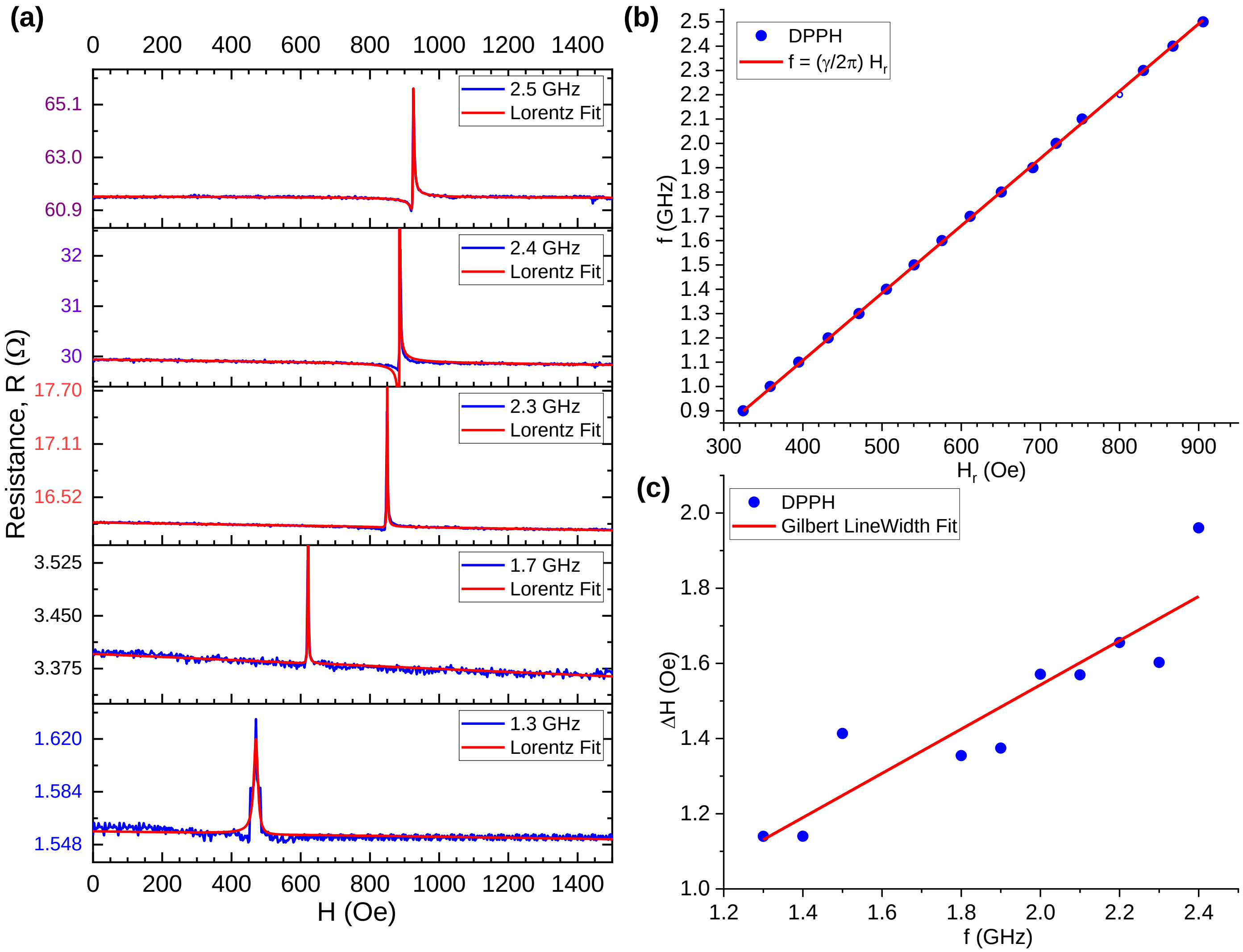

**Figure 3.** (a) Data (blue line) and Fit (red line) to a set of Lorentzian function to the $R(H)$ at 5 different frequencies. (b) Plot of the frequency $f$ vs $H_r$ (solid circles) and linear fit (red line) illustrating the linear relationship between $f$ and $H_r$ in EPR. (c) Plot of line width $\Delta H$ vs $f$ (solid circles) and the solid line illustrating the Gilbert linewidth fitting. The linewidth of our sample is ranging from 1.1 to 2.0 Oe, within reported values.

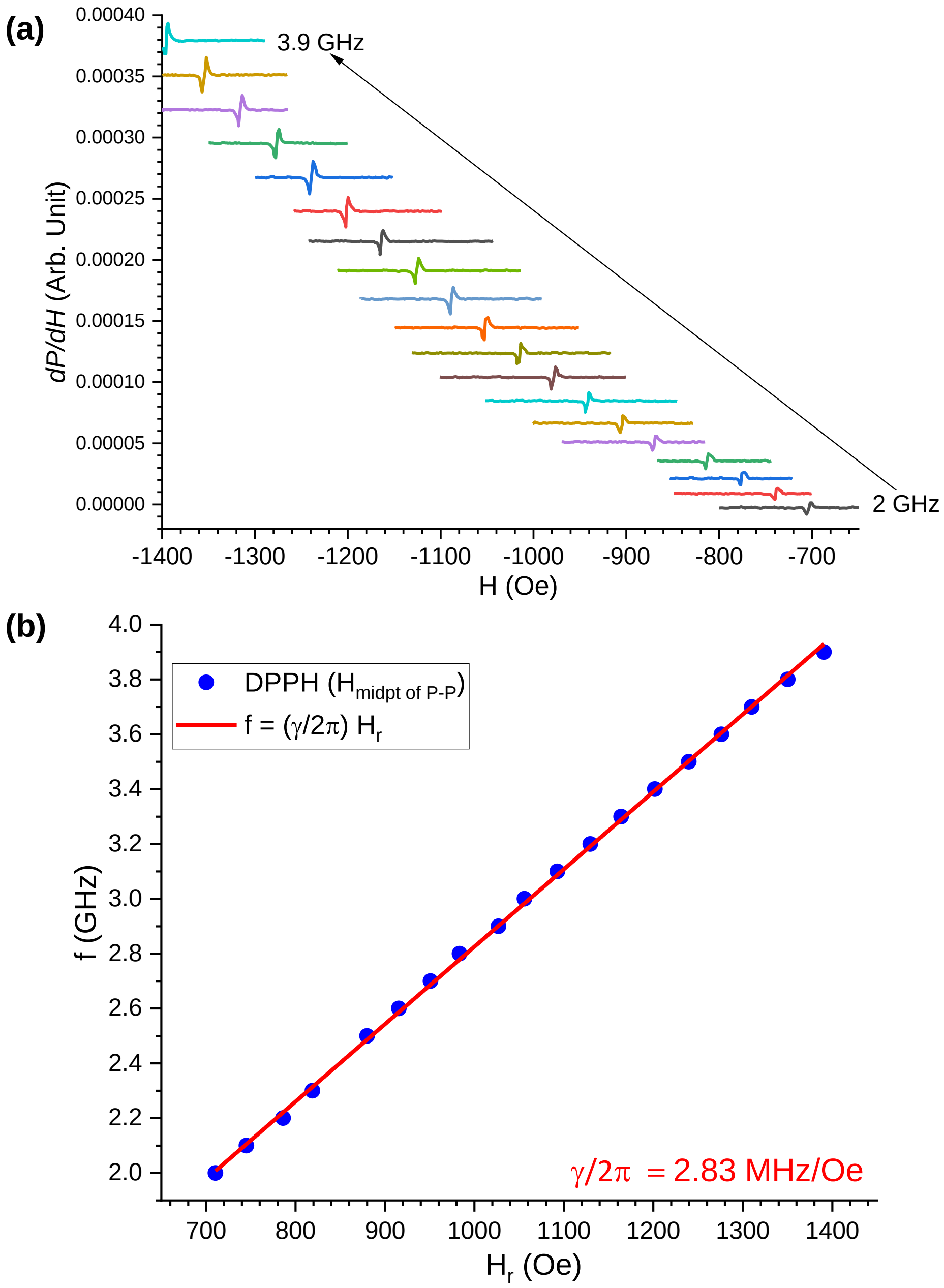

**Figure 4.** (a) The EPR spectroscopic signal (*dP/dH*) for the DPPH sample measured using the broad band spectrometer for excitation frequencies between 2 to 3.9 GHz. The signal is observed to shift towards the higher *H* as the frequencies increases. (b) Plot of *f* vs $H_r$ (solid circles) obtained using the *dP/dH* data. The red solid line represents the linear relationship indicating EPR.